\documentclass[12pt]{article}
\usepackage {graphicx}
\usepackage{makeidx}
\usepackage{multicol}
\usepackage{amssymb}
\usepackage{cite}
\input epsf

\begin{document}

\title{The mystery of two straight lines\\ in bacterial genome
statistics. Release 2007}

\author{
Alexander Gorban\thanks{ag153@le.ac.uk} \\ University of
Leicester, UK \\
 \\ Andrey Zinovyev\thanks{Andrei.Zinovyev@curie.fr} \\ Institut Curie,
Paris, France \\ and Institut des Hautes \'Etudes Scientifiques,
\\ Bures-sur-Yvette, France }

\date{}

\maketitle

\begin{abstract}
In special coordinates (codon position--specific nucleotide
frequencies) bacterial genomes form two straight lines in
9-dimensional space: one line for eubacterial genomes, another for
archaeal genomes. All the 348 distinct bacterial genomes available
in Genbank in April 2007, belong to these lines with high
accuracy. The main challenge now is to explain the observed high
accuracy.  The new phenomenon of complementary symmetry for codon
position--specific nucleotide frequencies is observed. The results
of analysis of several codon usage models are presented. We
demonstrate that the mean--field approximation, which is also
known as context--free, or complete independence model, or Segre
variety, can serve as a reasonable approximation to the real codon
usage. The first two principal components of codon usage correlate
strongly with genomic G+C content and the optimal growth
temperature respectively. The variation of codon usage along the
third component is related to the curvature of the mean-field
approximation. First three eigenvalues in codon usage PCA explain
59.1\%, 7.8\% and 4.7\% of variation. The eubacterial and archaeal
genomes codon usage is clearly distributed along two third order
curves with genomic G+C content as a parameter.
\end{abstract}

\section{Introduction}

The DNA double helix consists of two strands: the leading strand
and the lagging strand. Pairs of complementary nucleotides, base
pairs AT and GC, have different binding energies, and GC bond is
stronger. For statistical analysis of genome, a linear monotonic
function of the energy of strands binding, the G+C content (the
share of GC base pairs) is used most often. Some time ago
\cite{Muto} it was observed that the G+C content of various parts
of the genome (protein genes, stable RNA genes, and spacers)
reveals a positive linear correlation with the G+C content of
their whole genomic DNA. For coding parts, the energy of strand
binding is higher than for non-coding regions. In the studies
\cite{Yer1,Yer2,Blossey2005} this fact was used for gene finding
in a slightly more sophisticated form of ``DNA melting
temperature" difference: genes melt under higher temperature than
intergenic regions. This approach takes into account not only the
strands binding energy, but the entropy too. It is not a miracle
that the energy of strands binding defines some properties of
genome, but sometimes it seems surprising that so many properties
are defined by this energy. Now it has been proven that G+C
content can have a dramatic effect on the codon bias and on the
amino acid composition of the encoded proteins
\cite{Singer,Lobry97}.

\begin{figure}
\centering{
 a)\includegraphics[width=68mm, height=60mm]{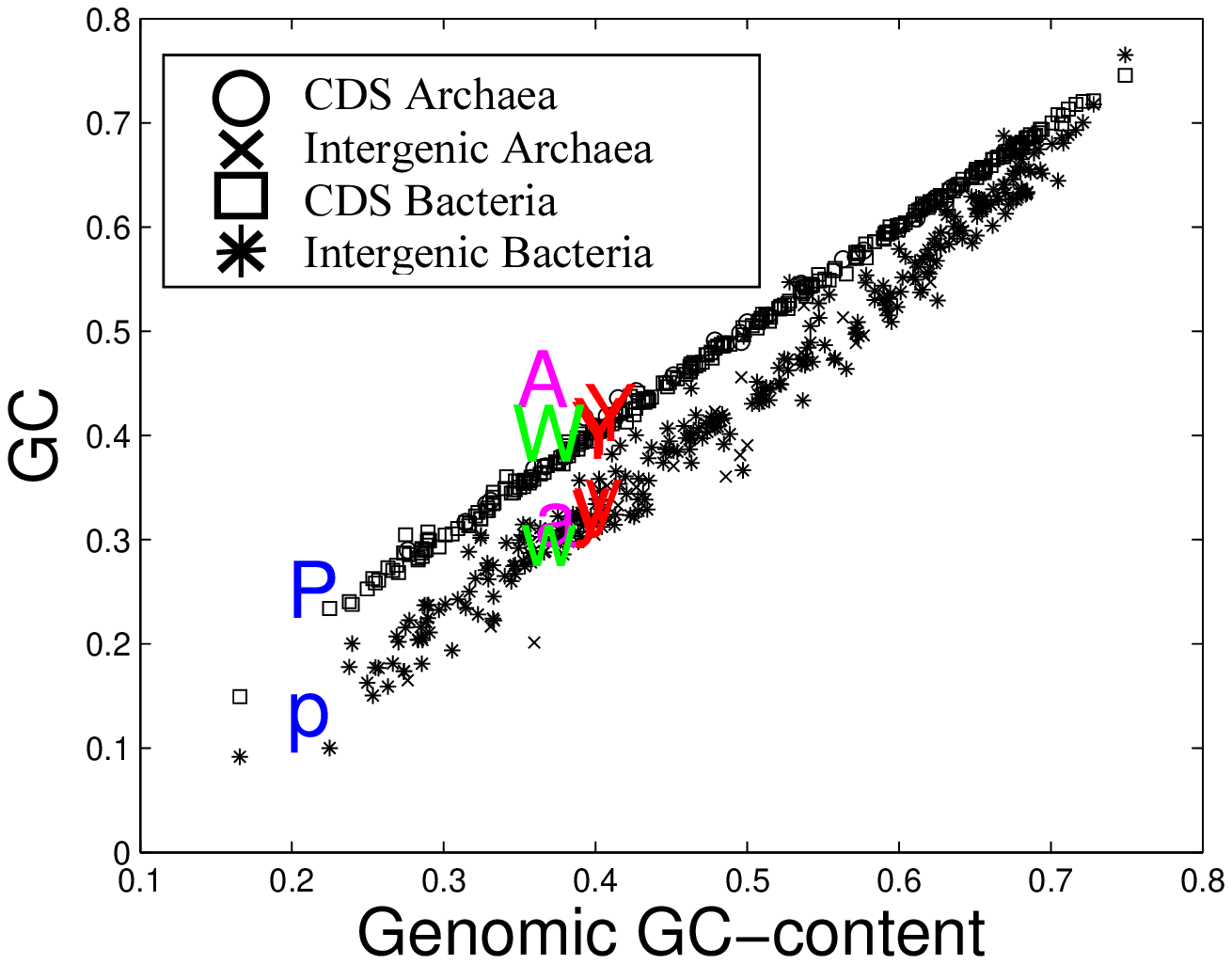}
 b)\includegraphics[width=68mm, height=60mm]{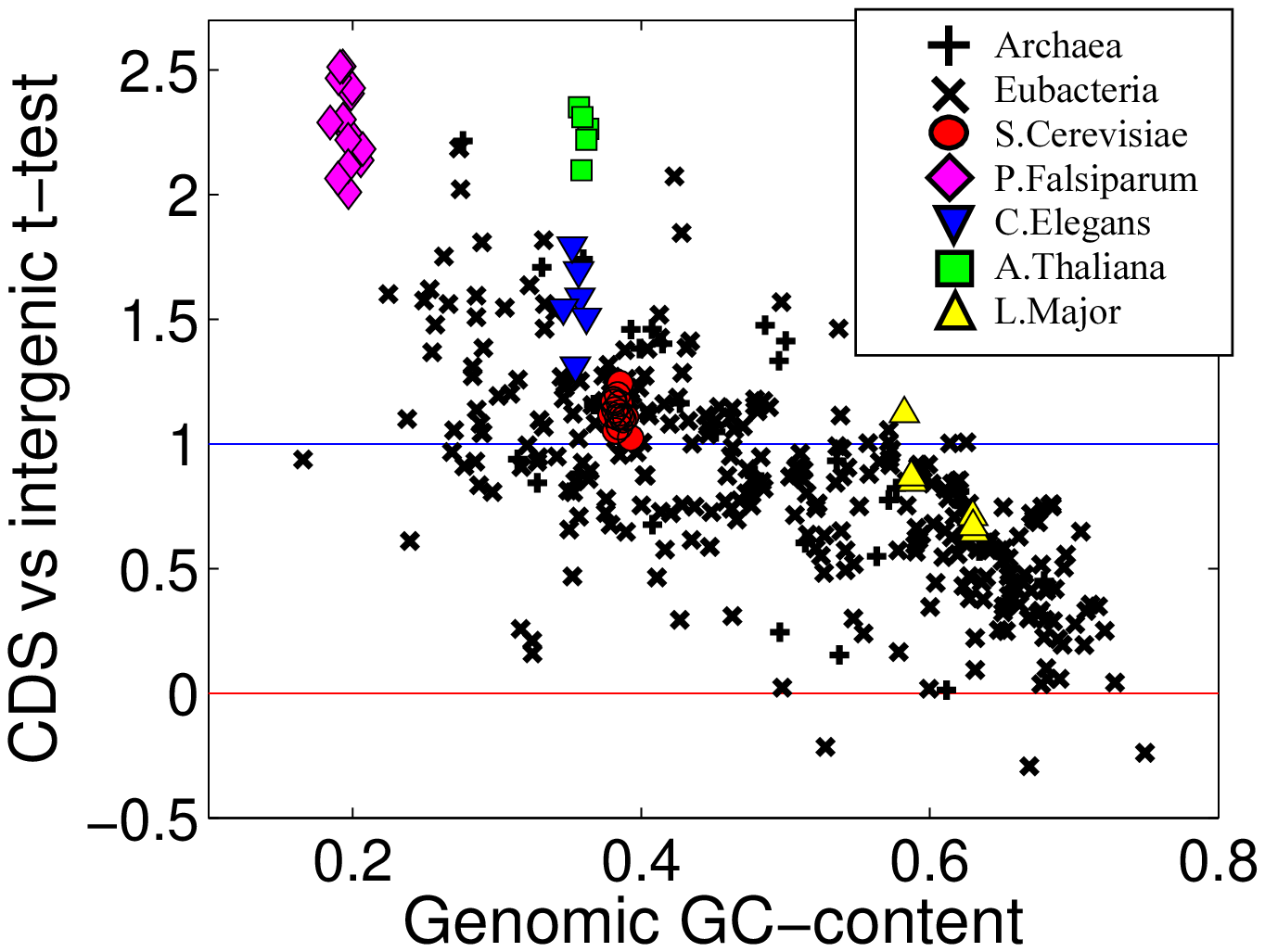}
 c)\includegraphics[width=68mm, height=60mm]{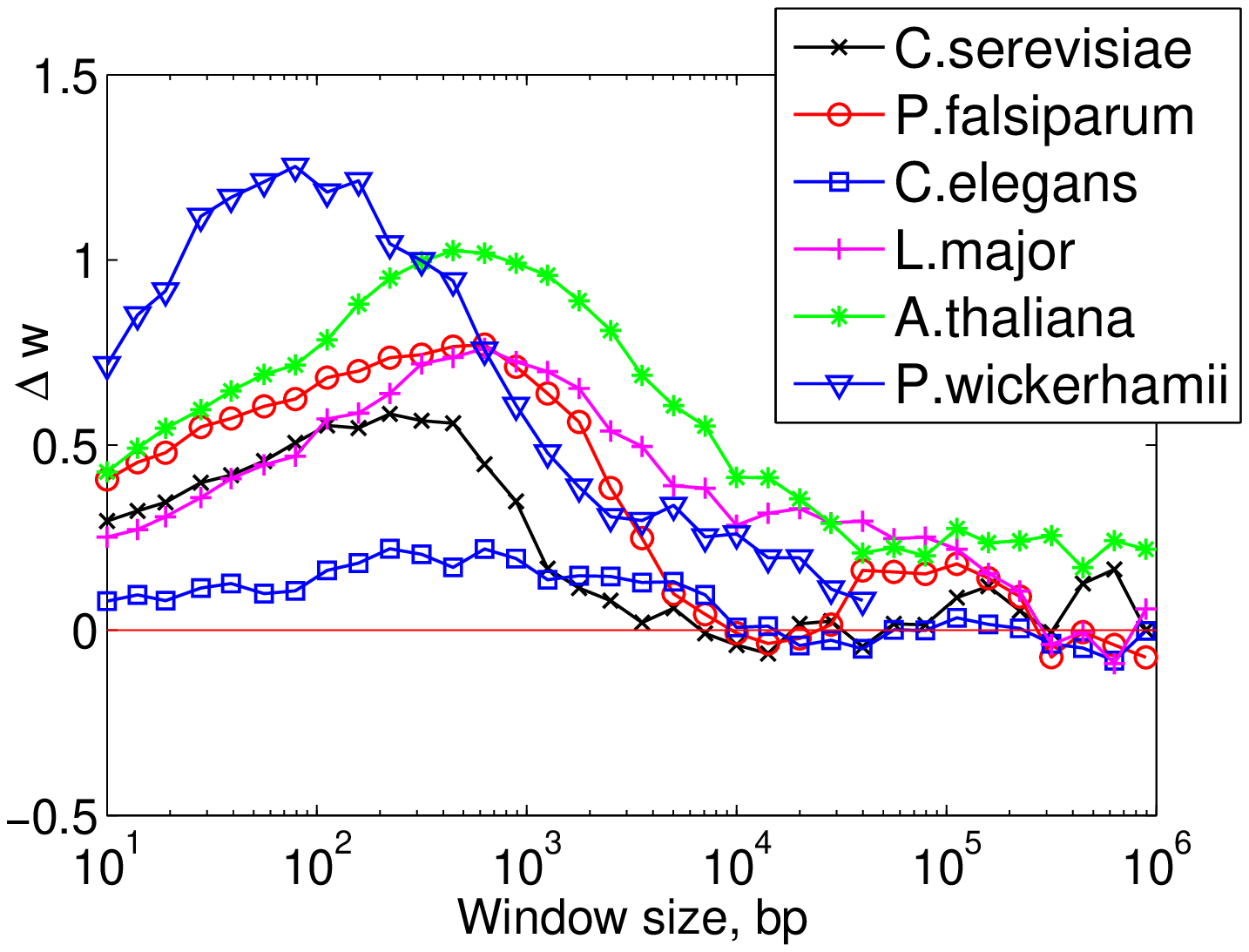}
 d)\includegraphics[width=68mm, height=60mm]{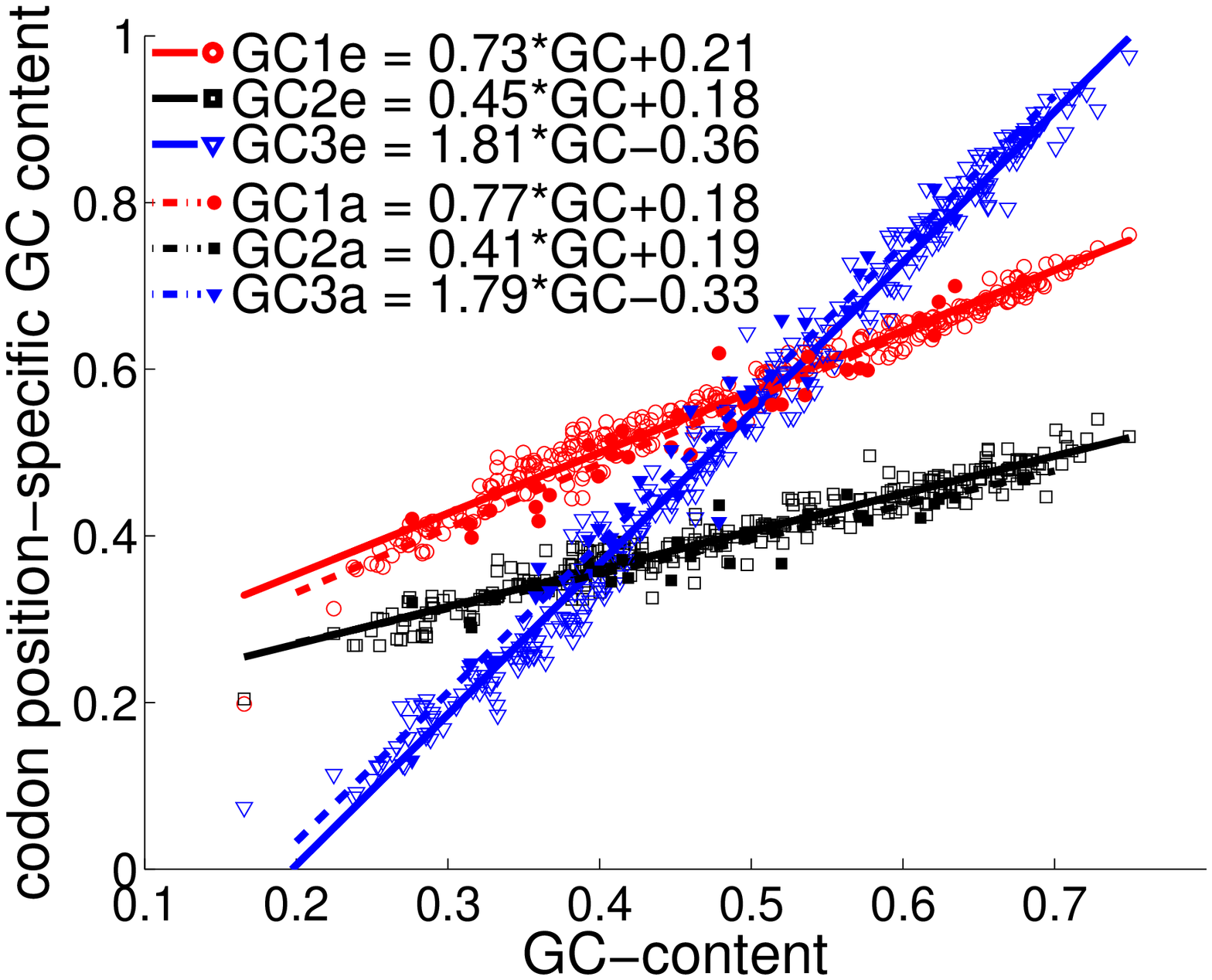}}
\caption{\label{gcgc} a) Average G+C content in coding and
intergenic regions as function of genomic G+C content. On the
graph 36 archaeal and 312 bacterial genomes are presented together
with several chromosomes of S.cerevisiae (represented by 'Y' and
'y' symbols, where capital 'Y' stands for G+C content in coding
regions and small 'y' stands for intergenic G+C content),
A.thaliana ('A' and 'a' symbols), C.elegans ('W' and 'w' symbols)
and P.falsiparum ('P' and 'p' symbols). As for S.cerevisiae,
capital letters stand for G+C content in coding regions, small
letters stand for intergenic G+C content. b) Difference between
average G+C content in coding and intergenic regions normalized to
unite variance (so-called t-test statistics for the hypothesis
GC$_{\rm coding}$ $>$ GC$_{\rm non-coding}$). c) Normalized
difference $\Delta_W$ between average G+C content in windows of
the length $W$ with the center belonging to coding and non-coding
regions. For $i$th base pair in the genomic sequence, the average
C+C content, GC$_W (i)$, in the window of length $W$ centered at
this base pair is calculated, GC$_{W {\rm cod} }$ and GC$_{W {\rm
non-cod} }$ are average values of GC$_W (i)$ for base pairs from
coding and non-coding regions, respectively, and $\Delta_W$ is the
normalized difference: $\Delta_W$=(GC$_{W {\rm cod }}$-GC$_{W {\rm
non-cod}}$)/(Var(GC$_{W {\rm cod }}$)+Var(GC$_{W {\rm non-cod
}}$))$^{1/2}$ (where Var is variance). d) Position--specific G+C
content, GC$_i$ as function of average G+C content, $(i=1,2,3)$.
Solid line and empty points correspond to 312 completed
eubacterial genomes, broken line and filled symbols correspond to
36 completed archaeal genomes.}
\end{figure}

The energy of strand binding in coding parts of genome is higher
than for non-coding regions, and this difference monotonically
decreases with genomic G+C content growth (Fig.~\ref{gcgc}a,b).
There are even two different effects: (i) the binding energy per
base pair in a coding region is higher than in nearest noncoding
regions and (ii) there exist regions in genome with typical length
$~10^5-10^6$ baispairs that have higher binding energy, and the
share of coding part in these G+C rich regions (i.e. concentration
of genes) is higher. These effects are illustrated by
Fig.~\ref{gcgc} (see also \cite{GorbanOpSys03}) for several
genomes.

There was a discussion, what should be chosen as the independent
variable for such plots based on a simple argument: both coding
and non-coding regions are parts of the genome, and it may be
better to use another part, GC3, average G+C content in the third
position of codons in coding part. Because of strong correlations
of GC$_i$ (the average G+C content in the $i$th position of codons
in coding part, $i=1,2,3$) (Fig.~\ref{gcgc}d), in this coordinate
we shall also obtain the same straight line, but with some
additional noise. We prefer not to discuss coordinate choice, and
follow the original work \cite{Muto} in data representation
because of the transparent physical sense of genomic G+C content
(it is monotonic linear function of energy density). Of course,
the genomic G+C content is a symmetric characteristic, it is the
same for both strands, and does not change under transposition
A$\leftrightarrow$T, G$\leftrightarrow$C. Hence, when studying
non-symmetric properties of strands, it may be useful to change
the coordinate choice \cite{LoSue}.

In this paper, we present results of analysis of the set of known
bacterial genomes (348 bacterial genomes from Genbank release April
2007). In some sense these results are experimental ones: we analyze
the experimental data from Genbank without additional ad hoc
theoretical hypothesis. In special coordinates (codon
position--specific nucleotide frequencies) bacterial genomes form
two straight lines in 9-dimensional space: one line for eubacterial
genomes, another for archaeal genomes. All the 348 bacterial genomes
belong to these lines with high accuracy, and these two lines are
certainly different. For lines in the codon usage space an
approximation of third order has appropriate accuracy, this is a
mean--field approximation (known also as complete independence model
or Segre variety \cite{ascb}).

The algebraic statistics paradigm \cite{ascb} works for this sort
of data. Simple straight lines or third order curves provide
unexpectedly accurate global models of data sets. We prove
universality and accuracy of this phenomenon. Some slices and
particular cases of these ``genome trajectories" are known
\cite{Borodovsky99}. They serve as important arguments in the
proof of the genome code universality \cite{Sueoka62}.

Of course, some of the linear functions appear because of simple
reasons. For example, the dependence CG$_{\rm cod}$(GC$_{\rm
genome}$) (Fig.~\ref{gcgc}a) should be linear with high accuracy
because of the  high proportion of coding part in bacterial
genomes (but even this dependency is more accurate than we could
prove on the basis of this proportion).

\section{Results}

\subsection{Two straight lines }

\begin{figure}

\centering{
\begin{tabular}{cc}

a)  \includegraphics[width=68mm, height=60mm]{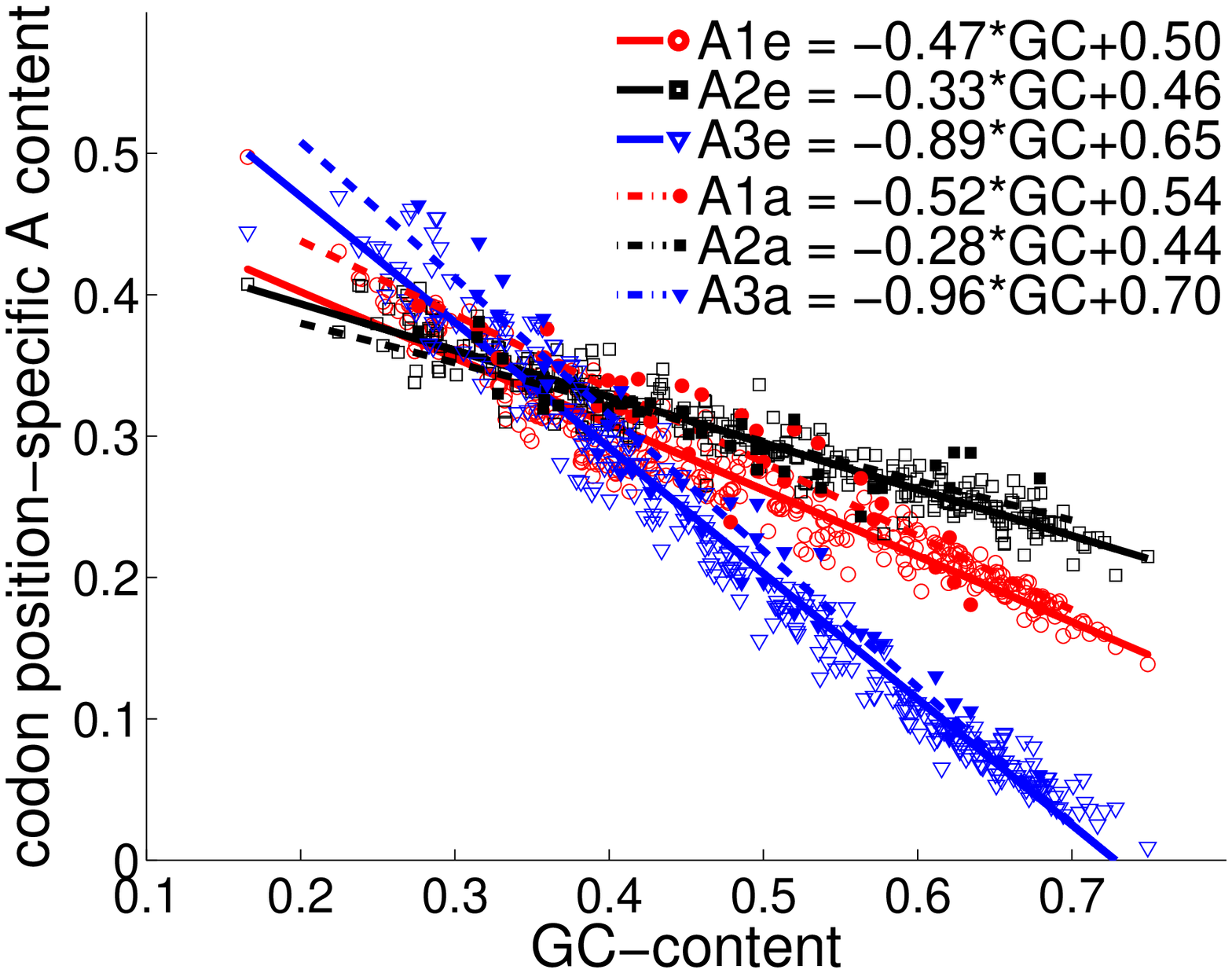}
 &
b) \includegraphics[width=68mm, height=60mm]{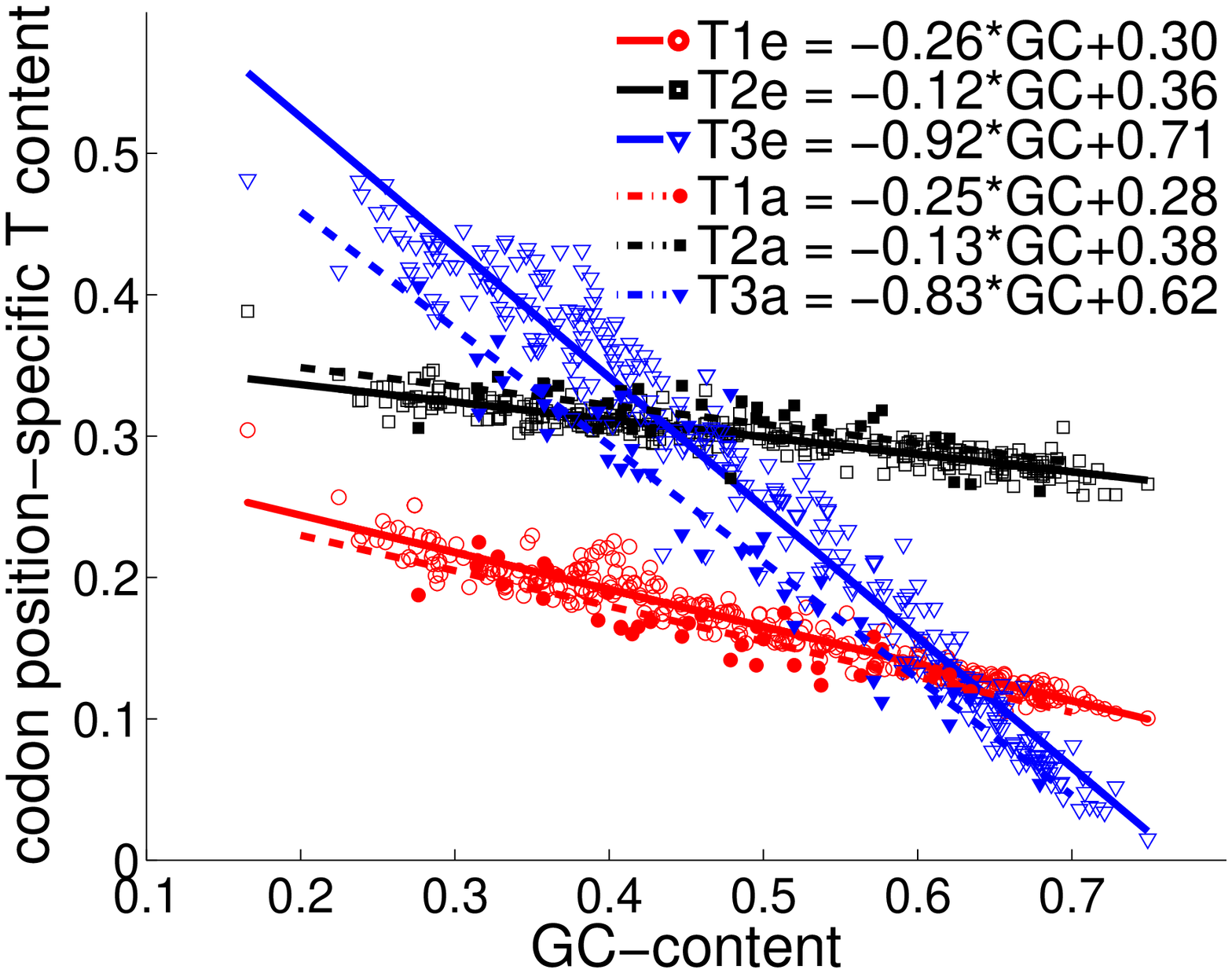}
  \\
c) \includegraphics[width=68mm, height=60mm]{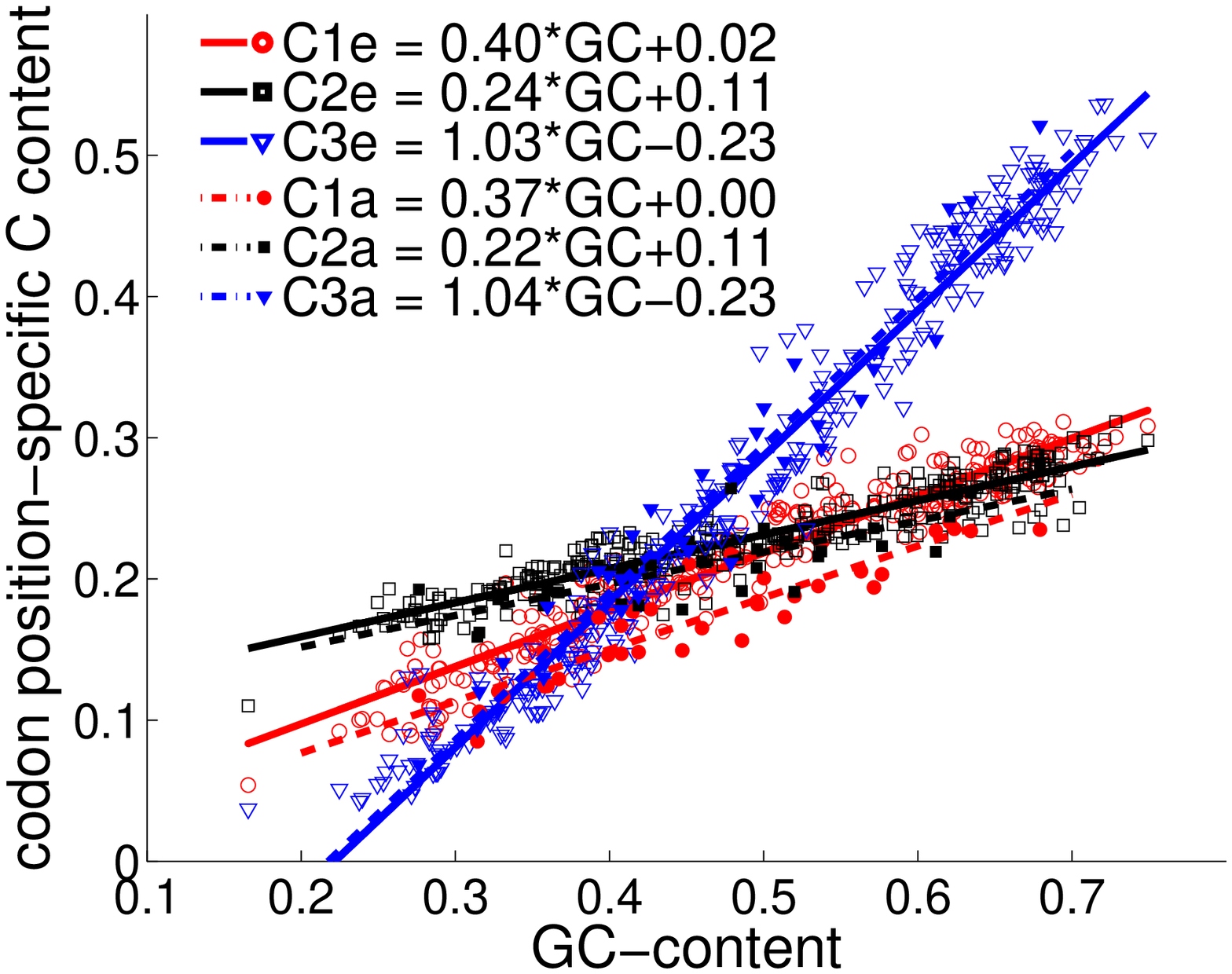}
 &
d) \includegraphics[width=68mm, height=60mm]{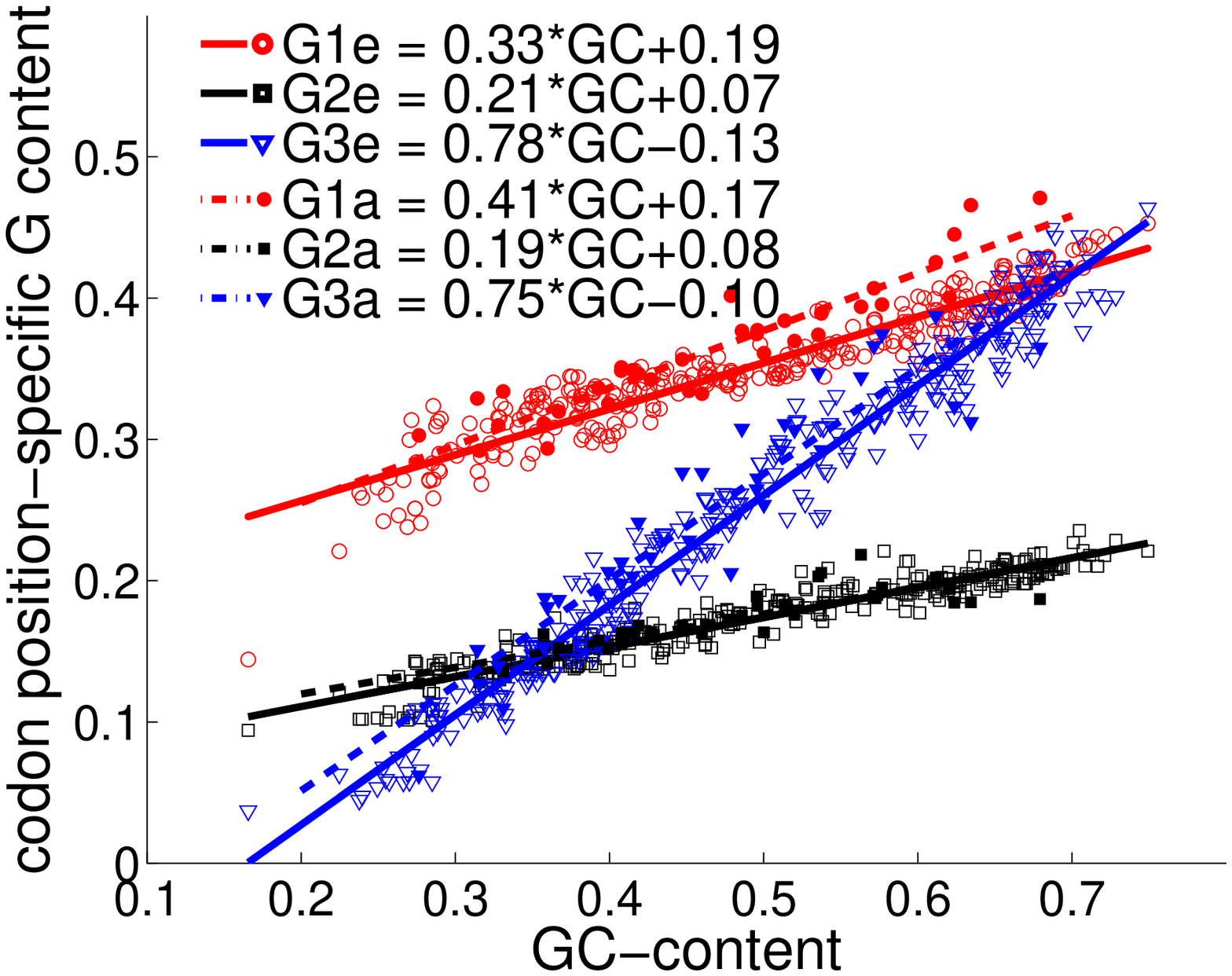}
\\
\end{tabular}
} \caption{\label{lines} Codon position--specific nucleotide
frequencies as functions of average G+C content. Solid line and
empty symbols correspond to 312 completed eubacterial genomes,
broken line and filled symbols correspond to 36 completed archaeal
genomes. }
\end{figure}

For each genome we can define the {\it codon position--specific
nucleotide frequencies} in the coding part of this genome. The
coding part of the genome is divided into codons. The codon
position--specific nucleotide frequencies are 12 numbers
$p_{\alpha}^i$, where $\alpha$ is the nucleotide symbol (A, C, G,
or T), and $i=1,2,3$ is the position number of nucleotide in a
codon. Among 12 frequencies $p_{\alpha}^i$ only 9 are independent
because of the normalization conditions $p_A^i + p_C^i + p_G^i +
p_T^i = 1 $. The possibility of non-uniqueness of separation of
the coding part into codons is here neglected. The vectors with
all possible coordinates $p_{\alpha}^i$ fill a 9-dimensional
polyhedron (a direct product of three tetrahedra). In this
9-dimensional space bacterial genomes form two straight lines: one
line for eubacterial genomes, another for archaeal genomes.  Both
of lines can be parametrized by genomic G+C content in a very
natural way, because $p_{\alpha}^i$ prove to be linear functions
of genomic G+C content with high accuracy. These functions are
different for different lines. The results of statistical analysis
(the regression lines together with experimental points) are
presented in Fig.~\ref{lines}. Significant differences between
eubacteria and arhaea are observed for $p_A^1$, $p_A^3$, $p_G^1$
and $p_T^3$ functions (see Table~\ref{coeffs} with intervals for
90\% confidence level). Available archaeal genomes are biased
towards thermophilic species and they are known to have their own
specific synonymous and non-synonymous codon usage \cite{Lobry03}.
This may be a possible reason for the observed difference between
linear dependencies (Fig.~\ref{lines}) for eubacterial and
archaeal genomes.

\begin{table}
\centering{ \caption{\label{coeffs}Coefficients for
$p_{\bullet}(GC)=k_1+k_2\times GC$ regression with intervals for
90\% confidence level. $R^2$ statistics shows how much variation
is explained by regression.} }\vspace{10pt} \small{
\begin{tabular}{l|r|r|r|r|r|r|}
Function&$k_1$&$k_2$&$k_1$ interval&$k_2$
interval&$R^2$&$p$-value\\ EUBACTERIA\\

A1&0.496&$-$0.467&(0.490;0.501)&($-$0.479;$-$0.456)&0.934&$<
10^{-16}$
\\ A2&0.459&$-$0.328&(0.455;0.464)&($-$0.337;$-$0.319)&0.919&$< 10^{-16}$
\\ A3&0.647&$-$0.889&(0.640;0.654)&($-$0.903;$-$0.875)&0.974&$< 10^{-16}$
\\ C1&0.016&0.404&(0.011;0.022)&(0.393;0.416)&0.916&$< 10^{-16}$ \\
C2&0.111&0.241&(0.106;0.115)&(0.232;0.250)&0.868&$< 10^{-16}$ \\
C3&$-$0.229&1.031&($-$0.237;$-$0.220)&(1.014;1.048)&0.970&$<
10^{-16}$
\\ G1&0.191&0.325&(0.186;0.197)&(0.315;0.336)&0.892&$< 10^{-16}$ \\
G2&0.069&0.210&(0.066;0.072)&(0.204;0.217)&0.904&$< 10^{-16}$ \\
G3&$-$0.128&0.777&($-$0.135;$-$0.122)&(0.765;0.790)&0.971&$<
10^{-16}$
\\ T1&0.297&$-$0.263&(0.293;0.300)&($-$0.269;$-$0.256)&0.930&$< 10^{-16}$
\\ T2&0.361&$-$0.123&(0.358;0.364)&($-$0.129;$-$0.118)&0.829&$< 10^{-16}$
\\ T3&0.710&$-$0.920&(0.701;0.718)&($-$0.937;$-$0.903)&0.961&$< 10^{-16}$
\\
GC1&0.208&0.730&(0.202;0.214)&(0.718;0.742)&0.970&$< 10^{-16}$ \\
GC2&0.180&0.451&(0.174;0.185)&(0.441;0.462)&0.941&$< 10^{-16}$ \\
GC3&$-$0.357&1.809&($-$0.367;$-$0.347)&(1.790;1.828)&0.988&$<
10^{-16}$
\\

ARCHAEA\\

A1&0.543&$-$0.522&(0.518;0.567)&($-$0.574;$-$0.470)&0.894&$<
10^{-16}$
\\
A2&0.436&$-$0.279&(0.415;0.457)&($-$0.323;$-$0.235)&0.769&2$\times
10^{-12}$
\\ A3&0.701&$-$0.964&(0.673;0.730)&($-$1.025;$-$0.904)&0.956&$< 10^{-16}$
\\ C1&0.003&0.366&($-$0.019;0.026)&(0.319;0.414)&0.833&8$\times 10^{-15}$ \\
C2&0.107&0.224&(0.085;0.129)&(0.177;0.271)&0.657&2$\times 10^{-9}$
\\ C3&$-$0.227&1.042&($-$0.262;$-$0.191)&(0.968;1.117)&0.943&$<
10^{-16}$
\\ G1&0.174&0.406&(0.153;0.195)&(0.361;0.451)&0.873&$10^{-16}$ \\
G2&0.082&0.189&(0.068;0.096)&(0.159;0.219)&0.770&2$\times
10^{-12}$
\\ G3&$-$0.098&0.748&($-$0.136;$-$0.060)&(0.667;0.828)&0.879&$<
10^{-16}$
\\ T1&0.280&$-$0.251&(0.262;0.298)&($-$0.288;$-$0.213)&0.791&4$\times
10^{-13}$
\\ T2&0.375&$-$0.134&(0.355;0.395)&($-$0.176;$-$0.092)&0.457&6$\times
10^{-6}$
\\ T3&0.624&$-$0.826&(0.589;0.659)&($-$0.900;$-$0.751)&0.912&$< 10^{-16}$
\\ GC1&0.177&0.773&(0.147;0.208)&(0.709;0.836)&0.925&$< 10^{-16}$ \\
GC2&0.189&0.413&(0.167;0.211)&(0.366;0.459)&0.870&$10^{-16}$
\\ GC3&$-$0.325&1.790&($-$0.368;$-$0.282)&(1.699;1.881)&0.970&$<
10^{-16}$
\\

\end{tabular}
}
\end{table}

\subsection{Complementary symmetry of codon
position--specific nucleotide frequencies}

There is one interesting feature of the observed dependencies on
Fig.~\ref{lines} which can be called {\it phenomenon of
complementary symmetry} in the region around $GC$=50\%. For given
position-specific frequencies $p^{(i)}_{\alpha}, i\in \{1,2,3\},
\alpha\in \{A,C,G,T\}$ let us define the measure of complementary
symmetry $S$:

\begin{equation}\label{symmmeas}
S :=
|p^{(1)}_A-p^{(3)}_T|+|p^{(2)}_A-p^{(2)}_T|+|p^{(3)}_A-p^{(1)}_T|+
|p^{(1)}_G-p^{(3)}_C|+|p^{(2)}_G-p^{(2)}_C|+|p^{(3)}_G-p^{(1)}_C|
\end{equation}

By its meaning, $S$ is half of $l_1$ distance from the
distribution of position-specific nucleotide frequencies to its
complementary distribution (as if it would be read from the
complementary strand such that $p_\alpha^{(i)}$ corresponds to
$p_{\bar{\alpha}}^{\bar{(i)}}$ where $\bar{A}=T$, $\bar{C}=G$,
$\bar{G}=C$, $\bar{T}=A$, $\bar{(1)}=(3)$, $\bar{(2)}=(2)$,
$\bar{(3)}=(1)$).

\begin{figure}[t]\centering{
\includegraphics[width=68mm, height=55mm]{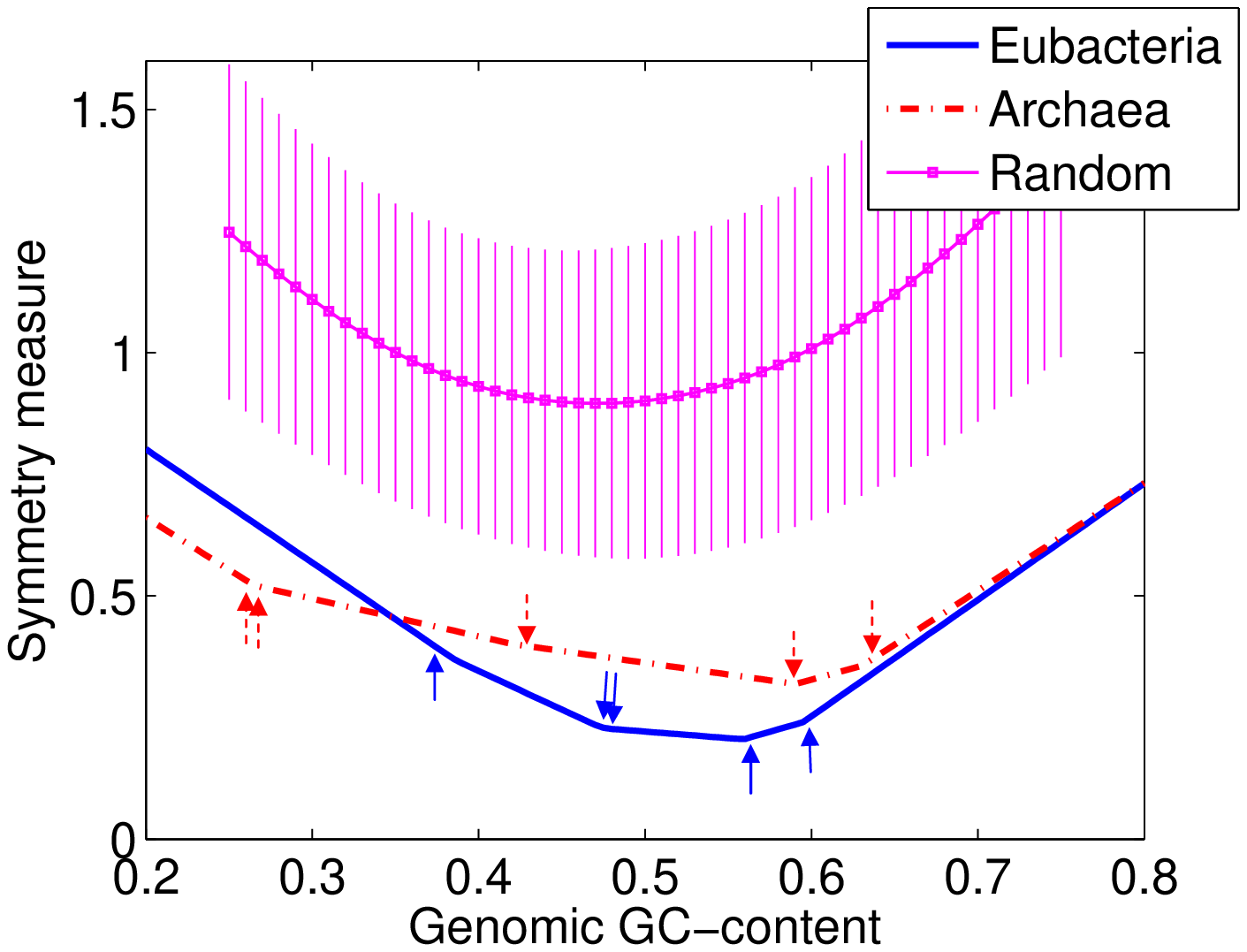}\hspace{5mm} \includegraphics[width=68mm,
height=55mm]{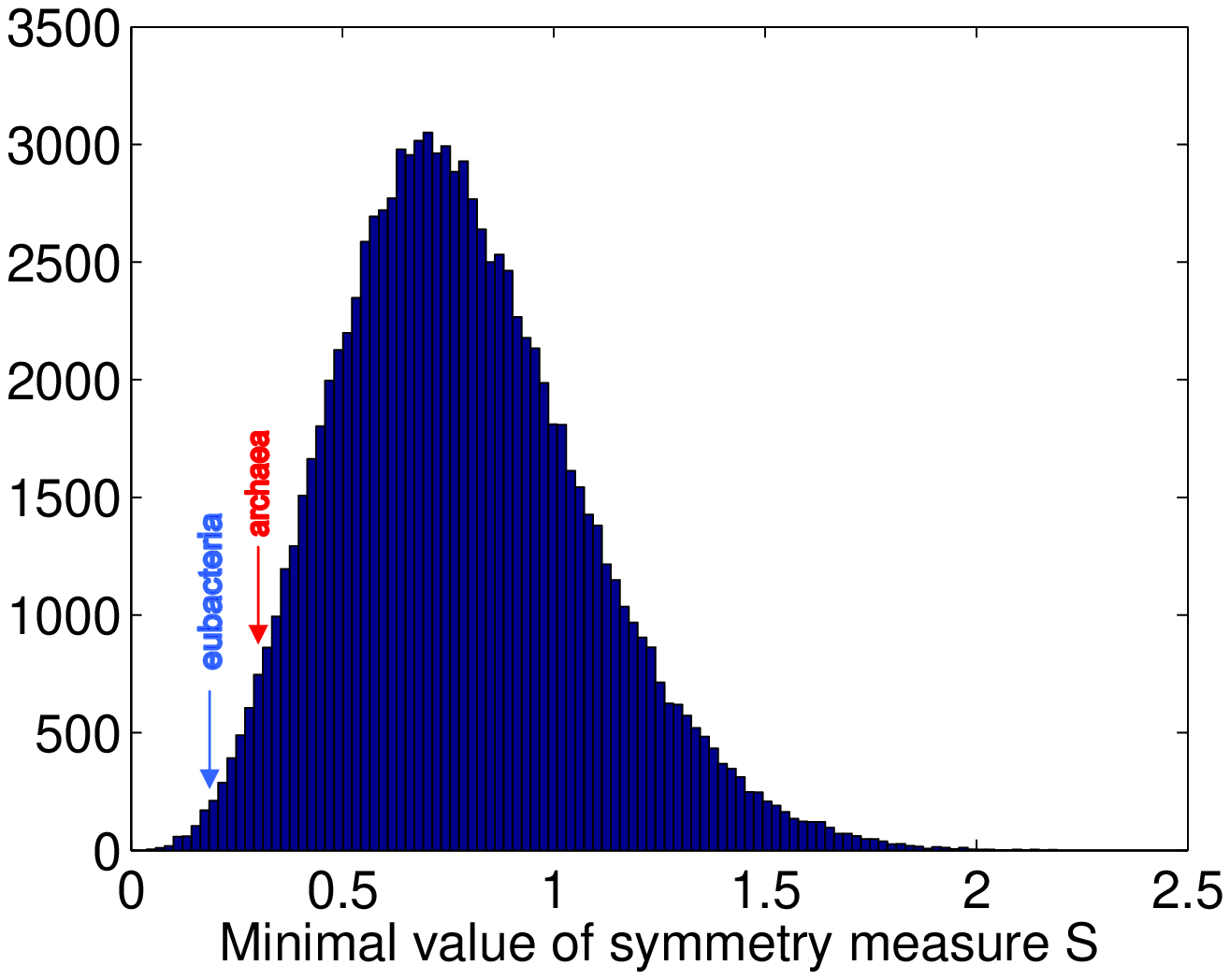}\\ \hspace{1cm}a)\hspace{4cm}b)
\caption{\label{symmetry} a) Complementary symmetry measure $S$
(\ref{symmmeas}) as a function of $GC$-content. Points where the
piece-wise linear curve changes its slope are shown by arrows. The
curve with square markers show the average behavior of $S$ values
in a random family of nucleotide frequency linear dependencies
$p(GC)=k_1+k_2\times GC$, together with standard deviations. b)
Distribution of minimal values of $S$ for the sampled family of
nucleotide frequency linear dependencies. The minimal values of
$S$ for eubacteria and archaea are shown by arrows.
 }}
\end{figure}

Let us now consider linear $p^{(i)}_{\alpha}(GC)$ dependencies
with parameters given in Table~\ref{coeffs}. Evidently, $S(GC)$
 is a piece-wise linear function, changing its slope every
time when any two ``complementary'' lines in Fig.~\ref{lines}
intersect. The behaviour of the function is shown at
Fig.~\ref{symmetry}. There are six possible slope breaks possible
(since we have six terms in (\ref{symmmeas})) but only five of
them happen with the positive $GC$-content: at $GC=38.6\%, 47.2\%,
47.8\%, 55.9\%, 59.4\%$ for eubacteria and $GC=26.4\%, 26.6\%,
42.0\%, 59.0\%, 63.0\%$ for archaea. Qualitatively, these breaks
divide $GC$ scale in several typical regions of different``codon
usage symmetry type'' discussed in \cite{7clulast,7clustPhysA}.

One can see from the Fig.~\ref{symmetry} that the minimal symmetry
measure $S_{\min}\approx 0.2$ is at $GC\approx 55.6\%$. Is the
value $0.2$ big or small? Or, what is the chance that $S_{\min}$
will have such value typically and should it be close to
$GC=50\%$? To answer this question, we considered a family of
linear dependencies with
$\{k_1^{p_\alpha^{(i)}},k_2^{p_\alpha^{(i)}}\}$ parameters. The
values of the parameters were limited by three requirements:

1) every linear dependency $p(GC) = k_1+k_2\times GC$ should give
feasible values of nucleotide frequencies (positive and less than
one) at the borders of $GC \in [0.25;0.75]$ interval;

2) the set of parameters should fulfill the normalization
requirements $\sum_{\alpha}p_{\alpha}^{(i)}(GC)=1$, i.e.
$\sum_{\alpha}k_1^{p_\alpha^{(i)}}=1$ and
$\sum_{\alpha}k_2^{p_\alpha^{(i)}}=0$;

3) the set of parameters should reproduce the observed
position-specific $GC$-content in bacteria, i.e.
$k_1^{p_C^{(i)}}+k_1^{p_G^{(i)}}=k_1^{p_{GC}^{(i)}}$,
$k_2^{p_C^{(i)}}+k_2^{p_G^{(i)}}=k_2^{p_{GC}^{(i)}}$, where
$k_1^{p_{GC}^{(i)}}$ and $k_2^{p_{GC}^{(i)}}$ are read from the
last three lines of the Table~\ref{coeffs}.

We randomly and uniformly sampled 100000 combinations of the
parameters in this family. The results are presented on
Fig.~\ref{symmetry}b. The expected value of $S_{\min}$ is 0.78 for
this family, and the probability of reaching $S_{\min}\leq 0.2$
(eubacteria) is $0.6\%$ and $S_{\min}\leq 0.32$ (archaea) is
$3.5\%$. The value of $GC$ corresponding to $S_{\min}$ is found
with probability $70\%$ in the $GC\in[0.35;0.60]$ interval where
it is distributed approximately uniformly (the data not shown). As
a conclusion, the lines on Fig.~\ref{lines} have rather particular
combination of parameters with respect to the minimal value of
complementary symmetry measure $S$.

Thus, there is another question we can ask about two lines: does
evolution tune the slope of the lines on Fig.~\ref{lines} such
that to obtain unusually symmetric pattern of codon frequencies in
the region $GC\approx 50\%$ or is it a historical trace of genome
evolution? What it might be needed for? This question is a
challenge for the future theoretical work.

\subsection{Validation of codon usage models}

\subsubsection{Mean--field (complete independence)
model and Segre embedding}

The average {\it codon usage} ($cu$) in genome is represented by
64 frequencies $p_{\alpha \beta \gamma}$, where $\alpha$, $\beta$,
and $\gamma$ are nucleotide symbols, $p_{\alpha \beta \gamma}$ is
the frequency of the codon $\alpha\beta\gamma$ in coding part of
genome. The {\it mean--field} approximation ($mf$) for codon usage
is $p_{\alpha \beta \gamma}^M =p_{\alpha}^1 p_{\beta}^2
p_{\gamma}^3$. This map $(p_{\alpha}^1, p_{\beta}^2, p_{\gamma}^3)
\mapsto p_{\alpha \beta \gamma}^M$ is the Segre embedding. It is
widely used in algebraic geometry to consider the cartesian
product of several projective spaces as a projective variety.

The Segre embedding transforms the straight lines
(Fig.~\ref{lines}) into lines of third order (two such lines: one
for eubacterial genomes, another for archaeal genomes) in the
63-dimensional simplex of codon usage (64 frequencies minus one
normalization condition). The codon usage for known genomes form
the clouds near these lines. These clouds have a distinctive
horseshoe form (see \cite{Lobry03} and Fig.~\ref{projections}). If
the mean--field approximation is applicable, then the clouds of
codon usage are close to these lines of third order.

Analysis of the mean--field approximation accuracy was, partially,
performed earlier in \cite{Knight}. The accuracy of mean--field
approximation in the 63-dimensional Euclidean space of codon usage
does not mean that it is accurate in all coordinate projections
uniformly. For example, it is worse for stop--codons. Violations
of the mean--field approximation for usage of amino acids Arg,
Val, Asp, Glu, Ser, and Cys were reported in \cite{Gautham}. In
this section we confirm some of these and other observations and
perform statistical evaluation of the mean-field codon usage model
for its ability to approximate real codon frequency distributions.

\subsubsection{Global approximation test}

Let us consider four statistical models of codon frequency
distribution $f_{ijk}$:
\begin{itemize}
\item{{\it Null background model} $f_{ijk} \approx p_{i}p_{j}p_{k}$,
where $p_{i}$ is simply a frequency of nucleotide $i$. This model
has $4-1=3$ parameters: 4 nucleotide frequencies $p_{i}$ minus one
linear relation between them, $\sum_i p_{i}=1$}
 \item{{\it Globally
average codon usage model} $f_{ijk} \approx {\hat{f}_{ijk}}$,
where ${\hat{f}_{ijk}} = 1/M \sum_{i=1..M}{f_{ijk}}$ is an average
codon frequency distribution observed in all $M$ available genomes
(some codons are more frequent than the others globally). This
model has $63=64-1$ (global, not genome dependent) parameters.}
 \item{{\it Mean-field or complete independence model} $f_{ijk} \approx
p^{1}_{i}p^{2}_{j}p^{3}_{k}$, where $p^{k}_{i}$ is the position
specific frequency of the $i$th nucleotide in the $k$th codon
position. This model has $12-3=9$ parameters: 12 position-specific
nucleotide frequencies (4 for each position) minus 3 linear
relations (1 for each position).}
 \item{{\it Partial independence
model}, discussed in \cite{Pachter2007} $f_{ijk} \approx
d_{ij}p^{3}_{k}$, where $d_{ij}$ is the frequency of the initial
dinucleotide {\it ij} in the codon {\it ijk} (18 parameters: this
model uses the full distribution of dinucleotides on the two first
positions that gives $16-1=15$ parameters, and nucleotide
frequencies on the third position give $4-1=3$ parameters).}
\end{itemize}

We have constructed all four models for a set of bacterial genomes
downloaded from GenBank, and tested the linear regression $f_{ijk}$
 {\it vs} $f^{model}_{ijk}$ for the totality of all observed frequencies in
348 genomes (there are $348\times 64 = 22272$ frequencies to
compare). The comparison of the models is given in
Table~\ref{ModelsGlobal}. It follows from the table that the
mean-field approximation is twice more precise with respect to the
null background distribution. The partial independence model,
having twice bigger number of parameters, brings 11\% of
improvement in comparison to the mean-field approximation. One can
conclude that the mean-field approximation can already serve as a
useful (although still very simple) model of real codon
frequencies.

\begin{table}
\centering{ \caption{\label{ModelsGlobal} Global comparison of the
statistical models of codon usage. $R^2$ statistics shows how much
variation is explained by regression. The estimations of the
regression $f_{ijk} = a+b\times f^{model}_{ijk}$ parameters are
given for $\alpha=10^{-3}$ confidence level.} }\vspace{10pt}
\begin{tabular}{|l|c|c|c|c}
Model name&\# parameters&a&b & $R^{2}$ \\ \hline Null background
model&3&(-0.0004;0.0007)&(0.9617;1.0241)&0.33 \\ Average codon
usage model&63(global)&(-0.0005;0.0005)&(0.9700;1.0300)&0.35 \\
Mean-field model&9&(-0.0001;0.0006)&(0.9675;1.0005)&0.633 \\
Partial independence model
&18&(-0.0001;0.0004)&(0.9756;1.0009)&0.748
\end{tabular}
\end{table}

\subsubsection{Approximation of individual genome codon distributions}

Now let us consider in some details how the mean-field
approximation works. First, let us consider some bacterial genome
$A$ for which we observe a codon frequency distribution
$f^{A}_{ijk}$ and we approximate it by the corresponding
mean-field model $m^{A}_{ijk}$. Let us also consider an
alternative global average codon usage model $\hat{f}_{ijk}$, i.e.
we also compare the given codon distribution to the global average
calculated for all available genomes. The question is: are all the
genomes approximated uniformly well? As in the previous section,
we evaluate the quality of the codon usage modeling by $R^2$ value
of the linear regression.

\begin{figure}\centering{
\includegraphics[width=92mm, height=68mm]{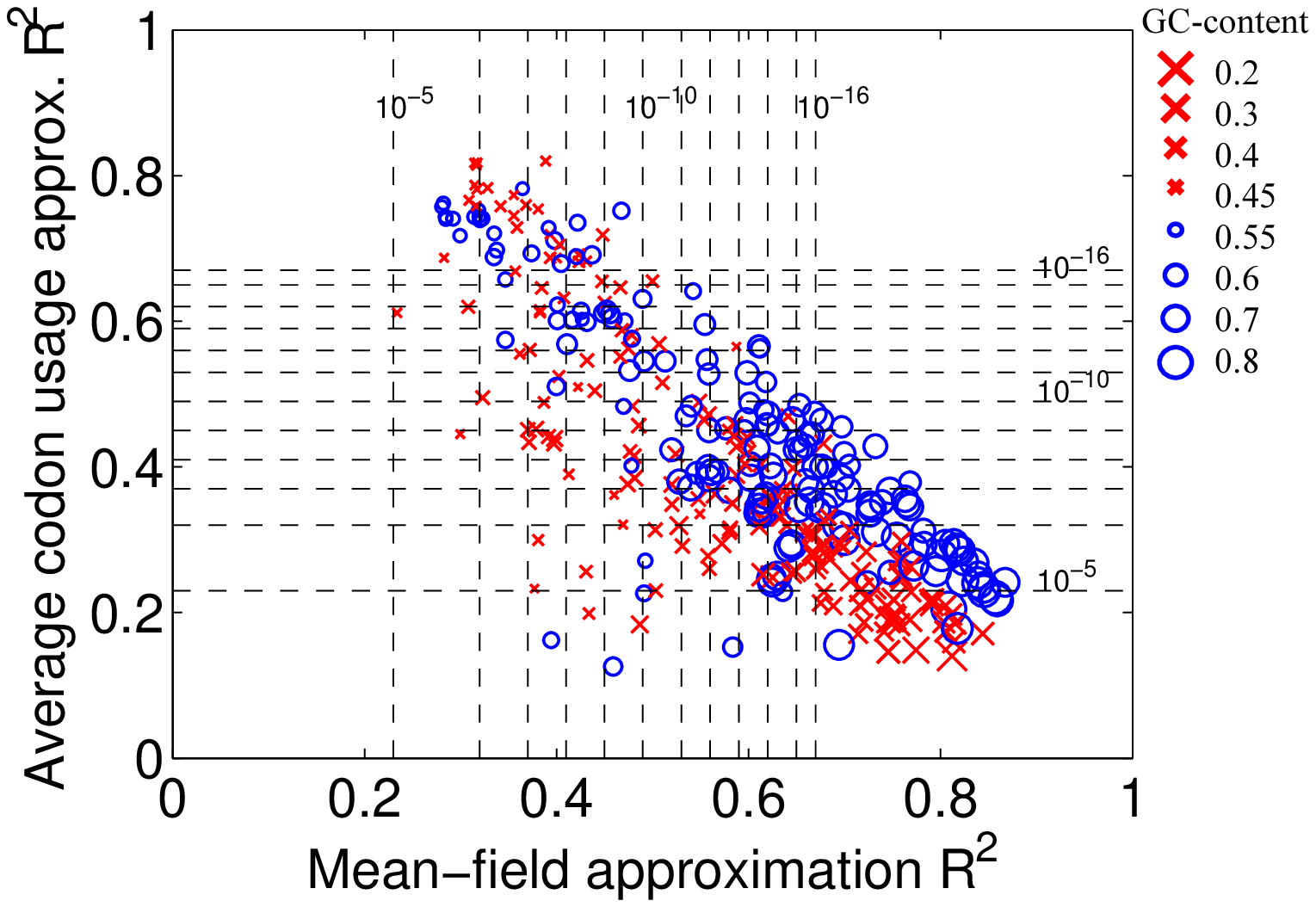}\\
\hspace{1cm}a)\\
\includegraphics[width=68mm, height=55mm]{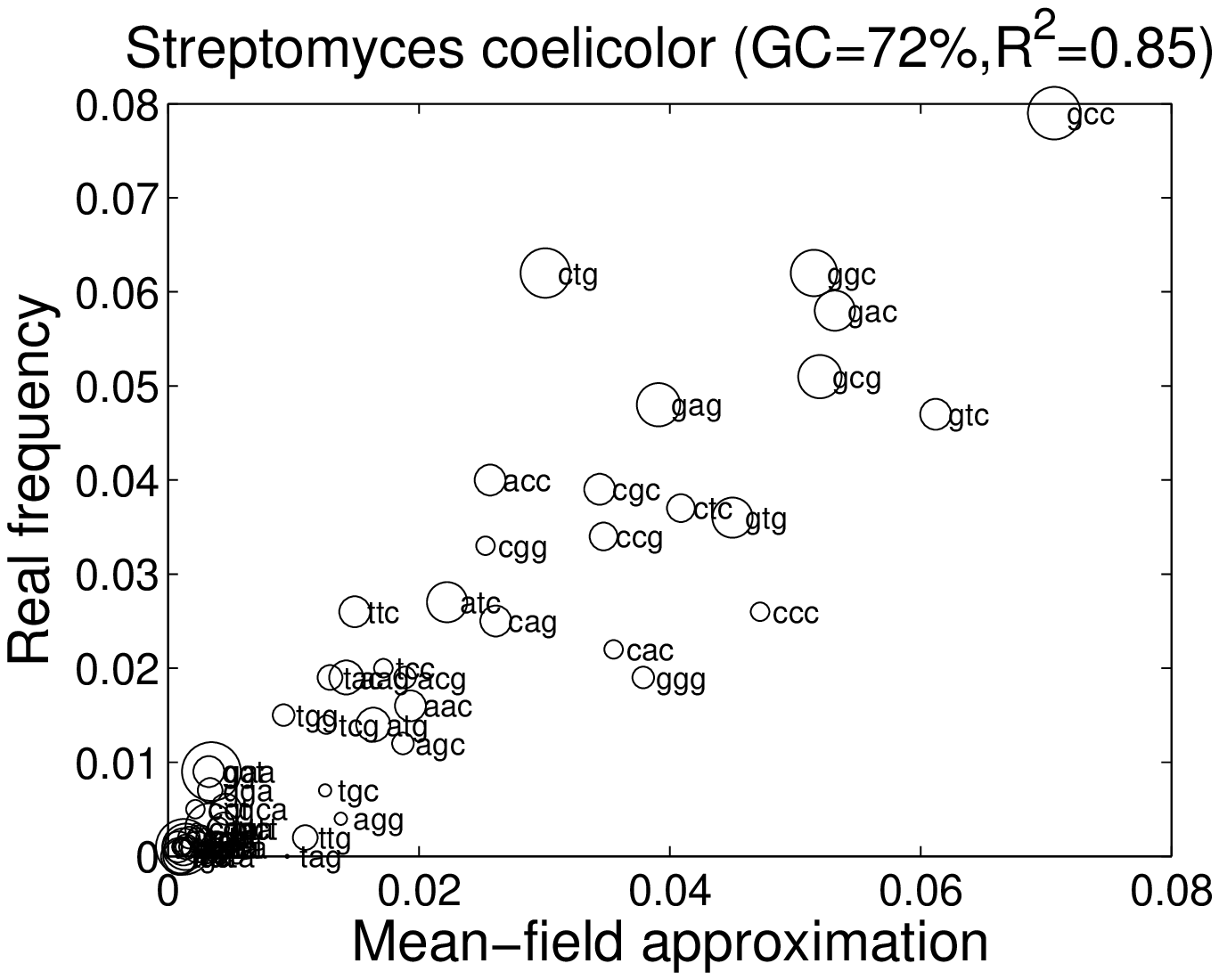}
\includegraphics[width=68mm, height=55mm]{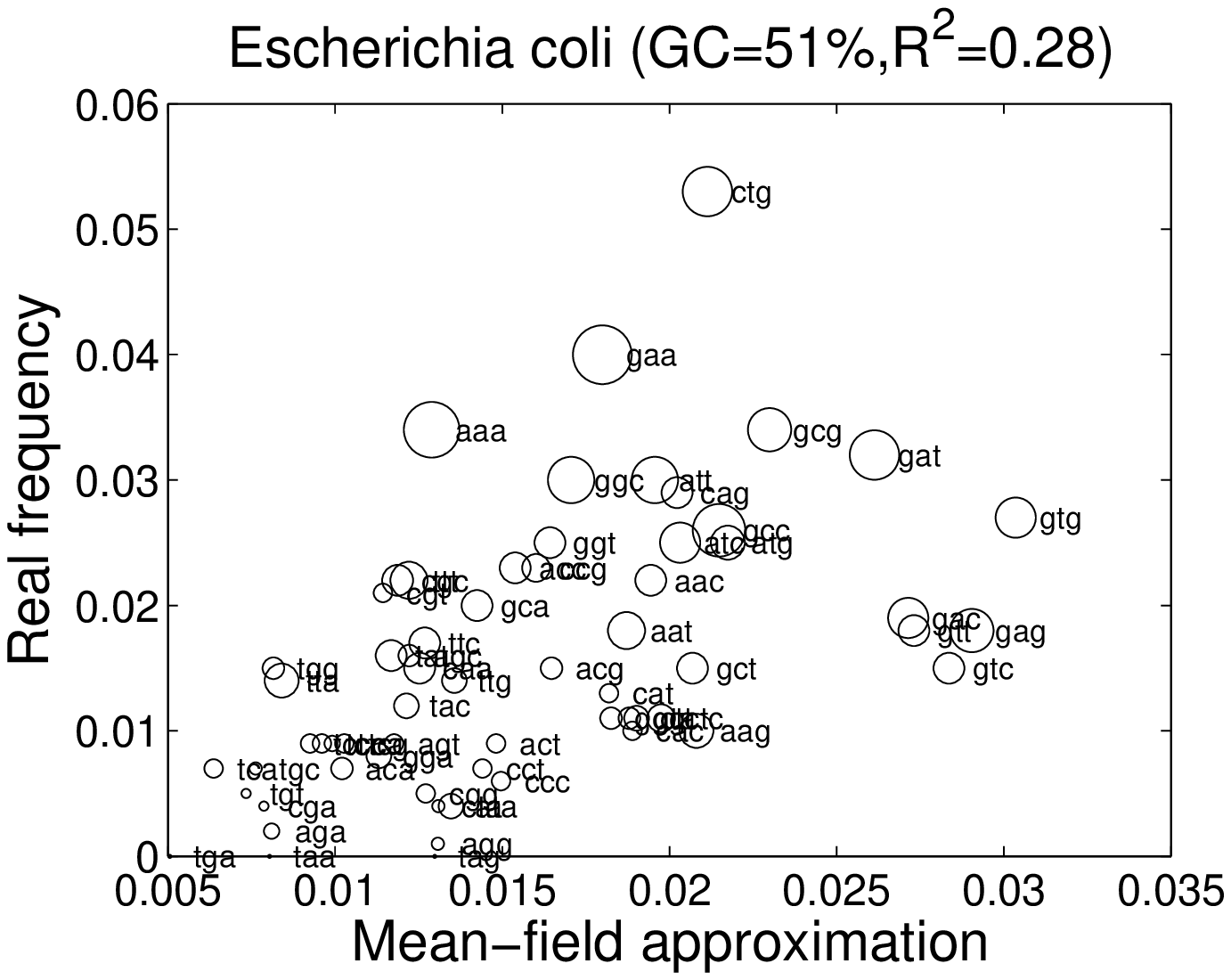}
\hspace{5cm}b)\hspace{5cm}c)
 \caption{\label{mf2hypot} a) $R^2$ values for
individual genomes for linear regression in two alternative codon
usage models: mean-field and global average codon usage. Every
point represent a genome. Circles are genomes with $GC$-content
$>50$\% and the crosses correspond to $GC$-content$<50$\%. Size of
the point is proportional to $|GC-50\%|$ value. Dashed lines
correspond to different powers of $p$-values. b,c) Approximation
of codon frequencies in the mean-field model for {\it
S.coelicolor} and {\it E.coli}. Each point corresponds to a codon.
The size of the point reflects global average frequency of the
codon in all bacterial genomes. $R^2$ value is given for the
mean-field model fit. It is clear that while the linear fit is
worse for {\it E.coli}, its codon frequencies are well correlated
with the global average codon usage $\hat{f}_{ijk}$.}}
\end{figure}

On the Fig.~\ref{mf2hypot}a we present comparison of two
alternative models $m_{ijk}$ and $\hat{f}_{ijk}$ for 348 available
genomes. The resulting conclusion is quite interesting: as a rule,
genomes with average GC-content close to 50\% are relatively poor
described by their mean-field approximations and the alternative
hypothesis $f^{A}_{ijk} = \hat{f}_{ijk}$ must be accepted. For
AT-rich and GC-rich genomes the mean-field approximation works
significantly better than $\hat{f}_{ijk}$. One of the explanation
of this observation is that close to {\it GC}=50\% a genome is
more free in choice of codons and this choice can be influenced by
the codon bias (by evolutionary pressure) which shapes the codon
distribution and can not be predicted from the position-specific
nucleotide frequencies (a genome selects the most optimal codon
among synonimous). However, there are genomes with {\it
GC}-content close to 50\% (like {\it A.marginale}, {\it
T.whipplei}, {\it H.walsbyi}) and whose codon usage can be well
estimated from the mean--field: this means that they are less
influenced by the codon bias.

There is a small group of genomes for which neither mean-field nor
average approximation work well (see Fig.~\ref{mf2hypot}a): the
most typical are {\it H.butylicus}, {\it A.pernix}, {\it
M.thermoautotrophicum}, {\it P.abyssi}, {\it P.torridus}, {\it
T.acidophilum}, {\it A.aeolicus}, {\it M.thermophila}, {\it
T.pendens}, {\it W.succinogenes}, {\it T.kodakaraensis}.
Interestingly, most of them are archaea habitating in extremal
conditions (4 of them belong to Thermoprotei, 2 to Thermococci, 2
to Thermoplasmata families, also one in Methanobacteria and one in
Methanomicrobia). Apparently, the environment of these bacteria
greatly influences their codon composition and makes them
different from the common trends.

On Fig.~\ref{mf2hypot}b two typical examples of genomes are
presented, one is well described by the mean-field approximation
and the other better fits the global average codon usage model.

\subsubsection{Approximation of individual codon frequencies}

\begin{figure}[t]\centering{
\includegraphics[width=80mm, height=60mm]{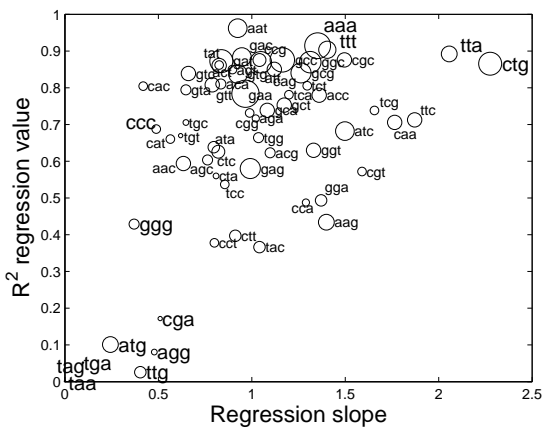}
 \caption{\label{CodonFig} a) $R^2$ values versus regression slope (parameter
 $b$ in
$f^{i}_{IJK} = a + b\times m^{i}_{IJK}$, where the regression is
estimated over all observed genomes $i=1..348$ for a fixed codon
{\it IJK}). Each point corresponds to a codon. The size of the point
reflects average frequency of the codon in all bacterial genomes.}}
\end{figure}

Now let us consider an individual codon $IKJ$ and look how its value
is approximated by the mean-field model through all 348 bacterial
genomes. In Fig.~\ref{CodonFig} the results of the linear
regressions are given for 64 codons. It is clear from the figure
that three usual stop codons {\it tag}, {\it tga}, {\it taa},
non-standard stop codon {\it agg}, standard start codon {\it atg}
and non-standard {\it ttg} as well as {\it cga} codon (which can be
easily mutated to {\it tga}) can not be well approximated by the
mean-field model. There is some tendency that more frequent codons
are better approximated (for example, relatively rare codons {\it
ccc} and {\it ggg} are much worse approximated than more frequent
{\it aaa} and {\it ttt}). Note that the regression slope in the fit
is not always close to 1 although it is distributed for various
codons around mean value 1 with standard deviation about 0.35. This
reflects the fact that along some directions of codon frequency
space, projection onto the mean--field manifold leads to
distortions, such as systematic under- or overestimated codon
frequencies. For example, the frequency of {\it ggg} codon is
systematically under-estimated after projection onto the
9-dimensional mean--field manifold while frequency of two
alternative start codons {\it ctg} and {\it tta} is systematically
over-estimated.

\subsection{PCA analysis of bacterial codon usage and the mean-field model}

\begin{figure}[t]
\includegraphics[width=60mm, height=40mm]{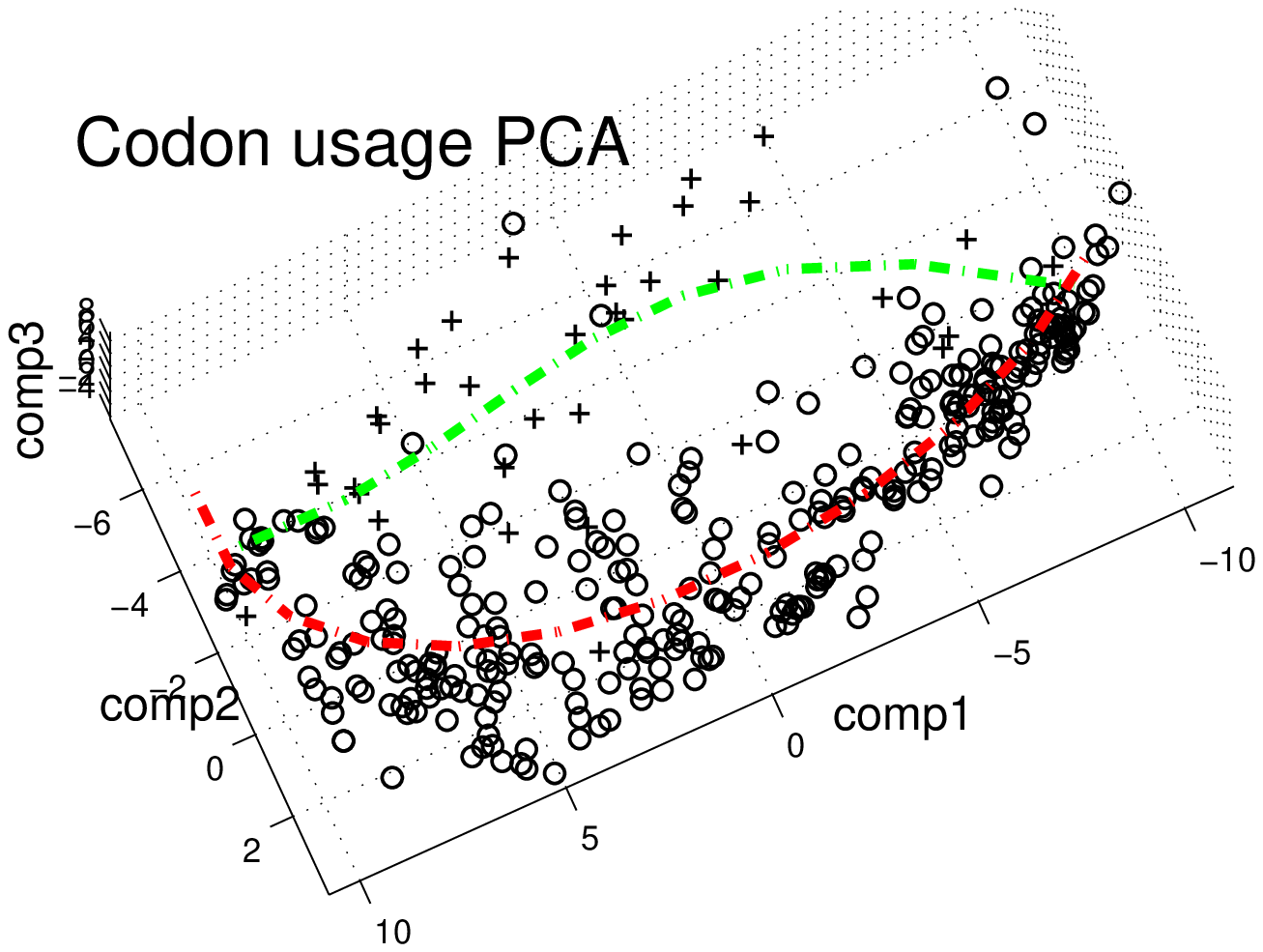}
\includegraphics[width=60mm, height=40mm]{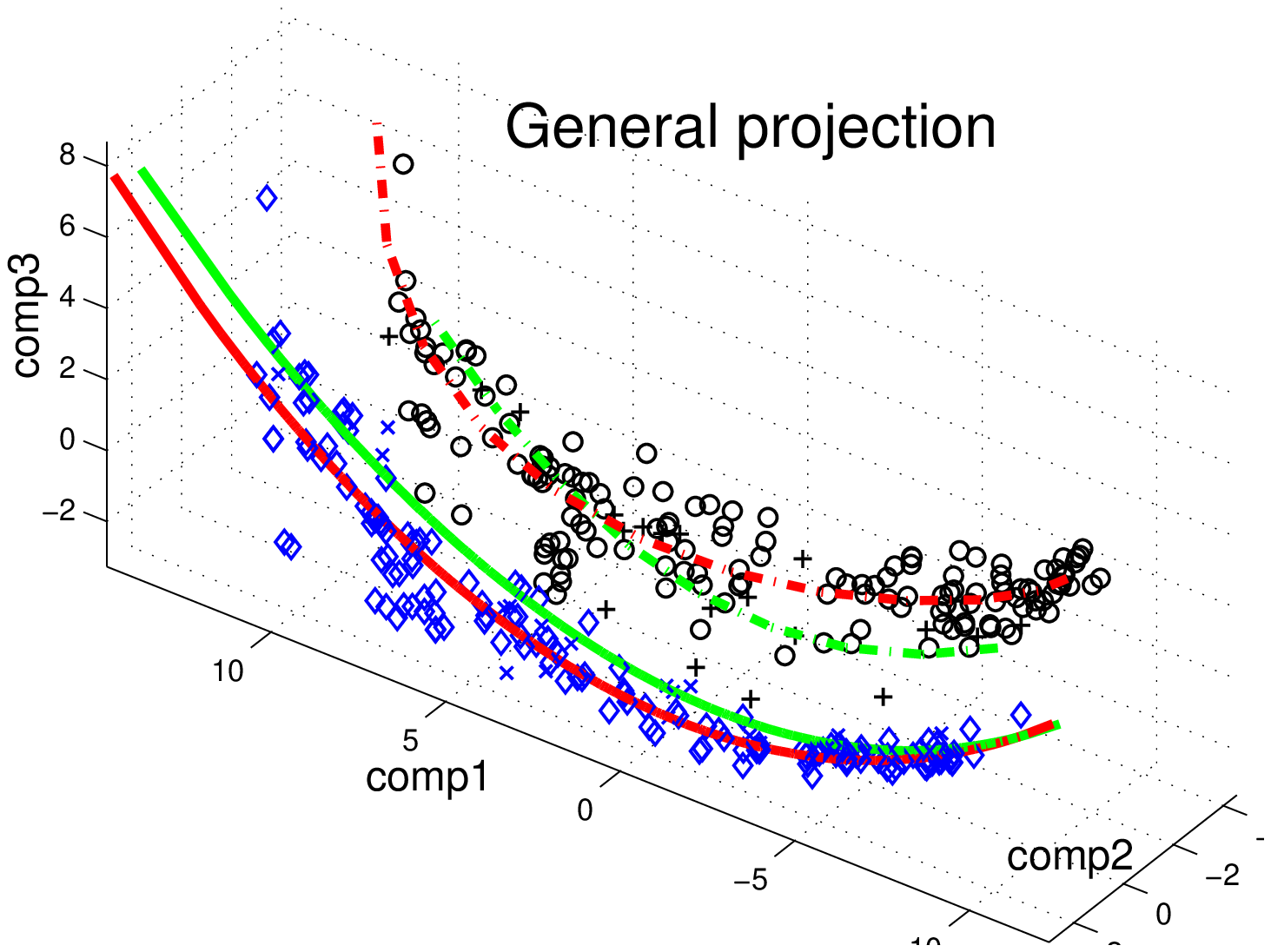}
\includegraphics[width=60mm, height=40mm]{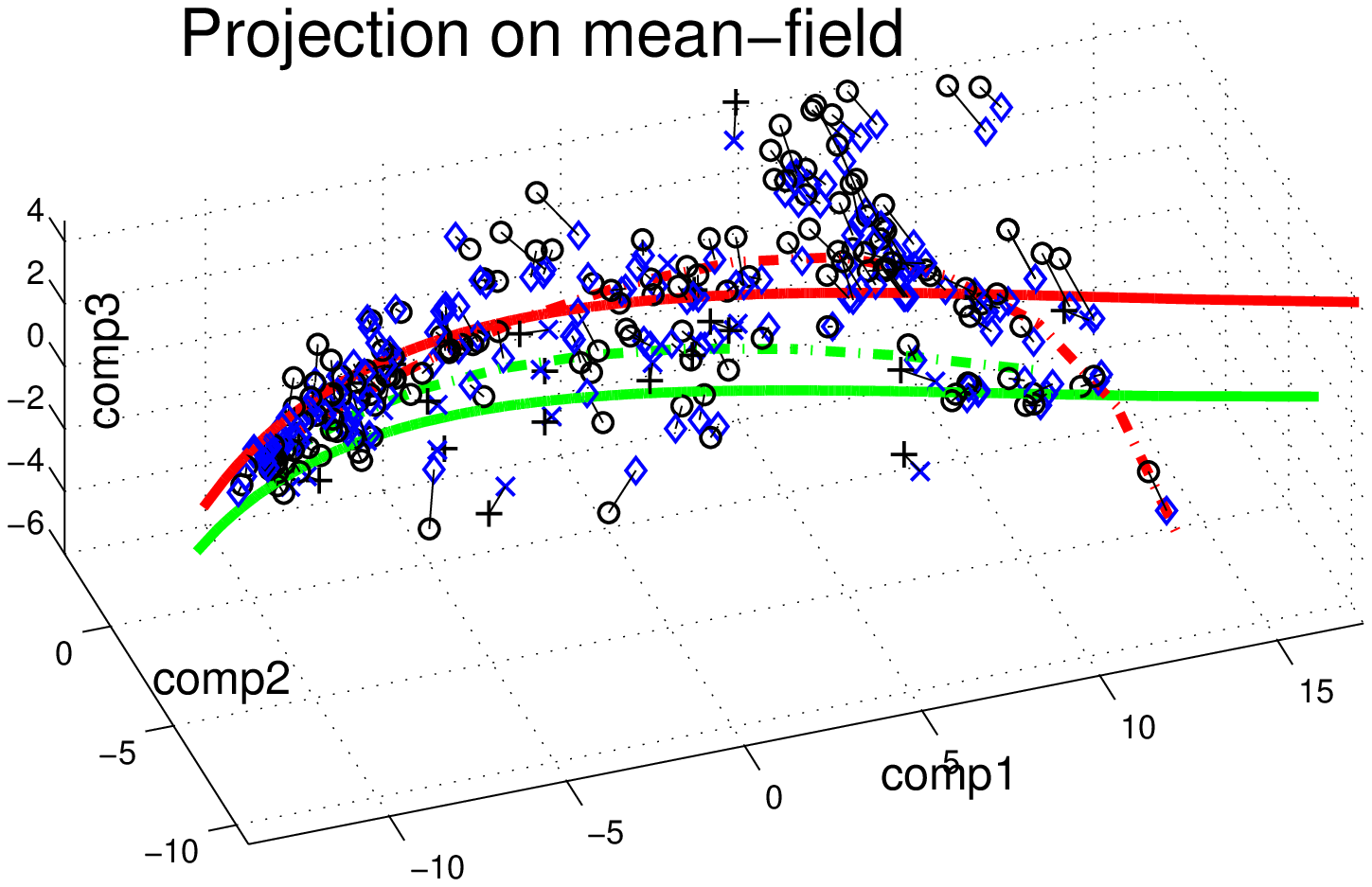}
\includegraphics[width=55mm, height=34mm]{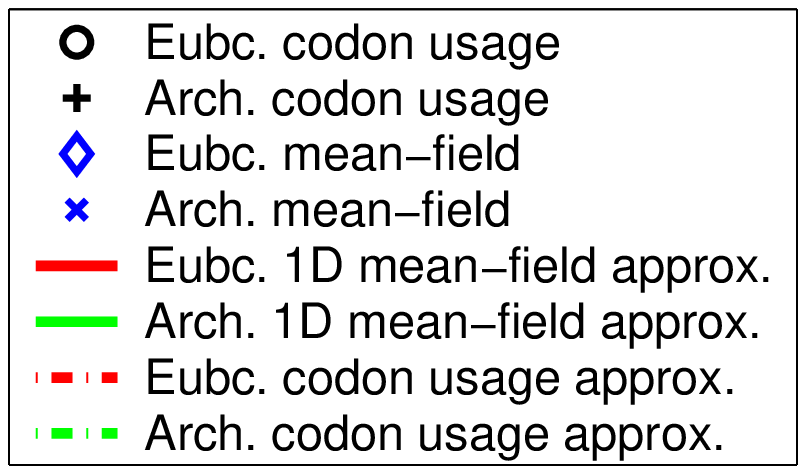}
\caption{\label{projections} PCA plots for the average codon usage
and the corresponding mean-field approximations for 348 bacterial
genomes; top-left: PCA plot of the average codon usage; top-right:
projection on the linear envelope of principal components
calculated for the average codon usage and the corresponding
mean-field approximations taken together; bottom-left: projection
on the linear envelope of principal components calculated for the
mean-field approximations only. Solid lines represent triplet
frequencies calculated with use of one-dimensional approximation
of mean-field approximation (see Fig.~\ref{lines}), stroked lines
represent fitting distribution of the average codon usage using
third order curves. On the projection onto mean-field
(bottom-left) same genomes represented by two points (codon usage
and its mean-field approximation) are shown connected. The lengths
of the lines correspond to the distance to the mean-field
manifold.}
\end{figure}

On Fig.~\ref{projections} we show 3D PCA plot\footnote{The basic
PCA procedure is quite simple. The $q$-dimensional PCA plane gives
the best approximation of data in the following sense. For a
finite sample, $\{x_i\}$ the mean point $y_0=\sum_i x_i /n$
minimizes the sum of Euclidean distance squares $\sum_i (x_i -
y_0)^2$. The first principal component vector, $y_1$, minimizes
the sum of distance squares from points $x_i$ to a straight line
$\{ y_0 + \alpha y_1 \ | \ \alpha \in R \}$, the second principal
component vector, $y_2$, is orthogonal to $y_1$ and minimizes the
sum of distance squares from points $f_i$ to a plane $\{ y_0 +
\alpha_1 y_1 + y_2 \psi_2 \ | \ \alpha_{1,2} \in R \}$, and so on.
Vectors of principal components $y_i$ are the eigenvectors of the
sample covariance matrix $\Sigma$. Presented PCA plot is the
projection onto 3D space spanned by the three first principal
component vectors. } visualizing genome average codon usage
distributions.

First three eigenvalues in codon usage PCA explain 59.1\%, 7.8\% and
4.7\% of variation. To estimate the number of principal components
we should retain (effective data dimension) we apply broken-stick
distribution test \cite{Cangelosi2007}. First values in the
broken-stick distribution for 64 dimensions are 7.3\%, 5.8\% and
5.0\%. This means that formally this test gives dimension 2,
however, the third values in both distributions are really close,
thus, it is reasonable to consider 3-dimensional PCA.

The first two principal components (coordinates in PCA basis) have
been shown (in \cite{Lynn}) to correlate strongly with genomic G+C
content and the optimal growth temperature respectively. The
variation of codon usage along the third component was not discussed
in the literature, but from Fig.~\ref{projections} it could be
extracted that it is related to the curvature of the mean-field
approximation manifold $\mbox{\boldmath{$M$}}$. The eubacterial and
archaeal genomes are clearly distributed along two trajectories and
we approximated them by fitting third order curves with genomic G+C
content as a parameter:

\begin{equation}\label{thirdOrderGCcurves}
p_{\alpha \beta \gamma} \approx a_{\alpha \beta
\gamma}GC^3+b_{\alpha \beta \gamma}GC^2+c_{\alpha \beta
\gamma}GC+d_{\alpha \beta \gamma},
\end{equation}

\noindent where coefficients $a_{\alpha \beta \gamma},b_{\alpha
\beta \gamma},c_{\alpha \beta \gamma},d_{\alpha \beta \gamma}$ are
fitted from data separately for eubacterial and archaeal genomes.
The order three is chosen because the mean-field approximations are
known to be distributed along trajectories of the third order
($p_{\alpha \beta \gamma}^M =p_{\alpha}^1(GC) p_{\beta}^2(GC)
p_{\gamma}^3(GC)$, where all dependencies on GC are close to linear,
see Fig.~\ref{lines}). These trajectories are also shown on
Fig.~\ref{projections}~(top-right, bottom-left). On
Fig.~\ref{projections}~(bottom-left) we use PCA to project both the
average codon usage $cu$ and the mean-field approximations $mf$ onto
the linear principal 3D manifold calculated for the mean-field
approximations only. In this projection\footnote{In fact, this 3D
manifold is almost perfectly embedded into the
$\mbox{\boldmath{$M$}}$ manifold. This means that this is a
principal ``view'' from within $\mbox{\boldmath{$M$}}$.} one can see
that, indeed, the mean-field approximation works quite nicely, but
on the other linear manifold, constructed for the totality of the
points, one can see that the mean-field approximations have a very
particular displacement (see Fig.~\ref{projections}~(top-right)).
This observation makes the story with mean-field approximation far
from being completely trivial: one has to explain why vectors
connecting $cu$ and $mf$ points are almost co-linear on this picture
and the $mf$ points are collected together: Is this simply an
artefact of projection or there is a specific direction of
information loss in the 64-dimensional space? At least it is clear
that the average codon usage is not simply randomly dispersed in the
vicinity of $\mbox{\boldmath{$M$}}$, but its ring-like spatial
structure is somehow specifically oriented relative to the
mean-field approximation manifold.

\section{Discussion}

We demonstrate that in  9-dimensional space of position--specific
nucleotide usage $p_{\alpha}^i$ bacterial genomes form two
straight lines: one line for eubacterial genomes, another for
archaeal genomes.  Both of lines can be parametrized by genomic
G+C content in a very natural way, because $p_{\alpha}^i$ prove to
be linear functions of genomic G+C content with high accuracy.
These functions are different for different lines. The new
phenomenon of complementary symmetry for codon position--specific
nucleotide frequencies is observed, which also needs theoretical
explanation. All  these observations are statistically verified
and do not depend on any codon usage model. These simple and
beautiful facts can serve as good benchmarks for testing the
mutation--selection--evolution models.

We show that the eubacterial and archaeal genomes in the
63-dimensional (64$-$1) space of codon usage are distributed along
two trajectories that are third order curves parametrized by
genomic G+C content, and in the 9-dimensional space (12$-$3) of
codon position--specific nucleotide frequencies bacterial genomes
form two straight lines: one line for eubacterial genomes, the
other for archaeal genomes. Some hints to observed structure were
reported recently in studies on multivariate analysis of bacterial
codon usage (for example, see Figure 6 from \cite{Lobry03}, or in
codon bias study in \cite{Carbone03}). Some other illustrations
can be found in our publications
\cite{mystery,7clulast,7clustPhysA} and on the web--site
\cite{Server}.

Our observations are consistent with previous studies (Sueoka's
neutrality plots, etc. \cite{Sueoka88,Wan}). Nevertheless, the
accuracy of the linear approximations (Fig.~\ref{lines}) seems to
be surprising. The correlation of amino-acids usage with genomic
G+C content was studied previously for 59 bacterial genomes
\cite{Lobry97}. It is not difficult now to build a
selection--mutation model that produces straight lines of
position--specific codon usage, just following the Sueoka theory
of directional mutation pressure and neutral molecular evolution
\cite{Sueoka88}. These lines appear as quasi-steady states of a
system of the first order reactions for a given concentration of
one component (the genomic G+C content). The main challenge is to
explain the observed high accuracy.

Codon usage is linked causally to a wide variety of both adaptive
and non-adaptive factors including tRNA abundance, gene expression
level, rates and patterns of mutations, protein structure, etc
\cite{Knight}. How to explain the accuracy of Fig.~\ref{lines} and
of Table~1 quantitatively? This challenging question already
stimulates new hypotheses and studies of algebraic structures in
genetic code and genomes \cite{FrapScar2006}, and new mutation
models are invented also \cite{Sciar05}.

There exists hierarchy of statistical models of codon usage
\cite{Pachter2007}: from the null background distribution (3
parameters), to the mean--field (or context--free or complete
independence model) approximation (9 parameters) and the partial
independence model (18 parameters). We tested these three models.
Of course, the highest model in the tested hierarchy, the partial
independence model, gives the best results, but the mean--field
model is only 11\% less precise (see Table~\ref{ModelsGlobal}) and
has twice less parameters. We can conclude that the mean-field
model can serve as a reasonable approximation to the real codon
usage. How to explain the accuracy of the simple mean--field
approximation (Table~\ref{ModelsGlobal} and
Fig.~\ref{projections}) quantitatively? We still do not know an
answer to this question.

{\bf Acknowledgement.} We are grateful to M. Gromov (IHES) and to
participants of workshop ``Geometry of Genome" (Leicester, August
2005) for stimulating discussion.

\end{document}